\documentclass[10pt]{article}
\begin{document}

\begin{center}\Large{\bf One-way speed of light?}\\

\normalsize\ J. Finkelstein\footnote[1]{
        Participating Guest, Lawrence Berkeley National Laboratory\\
        \hspace*{\parindent}\hspace*{.5em}
        Electronic address: JLFINKELSTEIN@lbl.gov}\\
        Department of Physics\\
        San Jos\'{e} State University\\San Jos\'{e}, CA 95192, U.S.A
\end{center}

\vspace{1cm}

In the October 2009 issue of the American Journal of Physics
 Greaves, Rodriguez and 
Ruiz-Camacho[1] report a measurement of           the one-way
speed of light.  This would seem to put them into conflict with the claim made
by, e.g.\ Reichenbach[2] that the one-way speed of light is to
some extent conventional because  its measurement depends on a
convention for the synchronization of distant clocks.  
Since the authors of this paper do not explicitly adopt any such
convention, the reader of this paper might wonder how, if Reichenbach
is correct, they are able to define, much less to measure, a one-way
speed.

In the experiment they describe, a light beam is sent from a laser to a
photosensor, and then the signal from the photosensor is transmitted
through a  coaxial cable back to the vicinity of the laser.  The
length of the cable is 23.73 m, and it is asserted that transmission
through the cable ``introduces a fixed time delay of 79 ns''.  The
authors point out that all timing is performed in a single 
place (the vicinity of the laser) so no convention for the
synchronization of distant clocks seems to be necessary.  However, the
assertion of a known time delay through the cable is only meaningful
if one imagines having synchronized clocks at the two ends of the
cable.  Therefore this assertion constitutes an implicit adoption of a
convention for distant synchronization.

What the experiment of Greaves, Rodriguez and Ruiz-Camacho actually
measures is the time for a round trip; the first leg of this round
trip is the light propagating from the laser to the photosensor, and
the second leg is the signal going through the coaxial cable from the
photosensor back to the vicinity of the laser.  It is the assumption
that the second leg is accomplished with a known speed (in particular,
the round-trip speed of light) that allows the speed of the first leg
to be determined.

\end{document}